%% file: main.tex
\documentclass{article}
\usepackage{spconf,amsmath,graphicx,hyperref}

\usepackage{multirow}
\usepackage{graphicx}
\usepackage{amssymb, amsmath, bm, mathtools,array}
\usepackage{textcomp}
\usepackage{hyperref}
\usepackage{verbatim,lipsum}
\usepackage{booktabs}
\usepackage{tcolorbox}
\usepackage{lscape}
\usepackage{subcaption}
\usepackage{multirow}
\usepackage{colortbl}
\usepackage{color}
\usepackage{multibib}

\RequirePackage{color}\definecolor{BLUE}{rgb}{0,0,0}

\newcommand{\mAASIST}{{\texttt{AASIST}}}
\newcommand{\mWVS}{{\texttt{W2V}}}
\newcommand{\mXLSR}{{\texttt{XSLR2b}}}

\newcommand{\symbolvar}[1]{\boldsymbol{\mathrm{#1}}}
\newcommand{\symbolvec}[1]{\boldsymbol{#1}}

\newcommand{\symboldomain}[1]{\mathbb{#1}}
\newcommand{\symbolset}[1]{\mathcal{#1}}
\newcommand{\symboltag}[1]{\texttt{#1}}

\title{Towards Data Drift Monitoring for Speech Deepfake Detection in the context of MLOps}
\name{Xin Wang\thanks{This study is supported by JST PRESTO Grant (JPMJPR23P9) and JST AIP Acceleration Research (JPMJCR24U3). This study was carried out on TSUBAME4.0 supercomputer at the Institute of Science Tokyo.}, Wanying Ge, Junichi Yamagishi}
\address{National Institute of Informatics, Japan}

\begin{document}
\ninept
\maketitle
\begin{abstract}
When being delivered in applications or services on the cloud, static speech deepfake detectors that are not updated will become vulnerable to newly created speech deepfake attacks. 
From the perspective of machine learning operations (MLOps), this paper tries to answer whether we can monitor new and unseen speech deepfake data that drifts away from a seen reference data set. We further ask, if drift is detected, whether we can fine-tune the detector using similarly drifted data, reduce the drift, and improve the detection performance.  
On a toy dataset and the large-scale MLAAD dataset, we show that the drift caused by new text-to-speech (TTS) attacks can be monitored using distances between the distributions of the new data and reference data. Furthermore, we demonstrate that fine-tuning the detector using data generated by the new TTS deepfakes can reduce the drift and the detection error rates. 
\end{abstract}

\begin{keywords}
Speech deepfake detection, concept drift, deep learning, biometric
\end{keywords}
\section{Introduction}
\label{sec:intro}

A speech deepfake detector takes an input waveform $\symbolvec{x}\in\symboldomain{R}^T$ and produces a label $y\in\{\text{fake}, \text{real}\}$ indicating whether the input is a deepfake or not. 
Our community has mainly approached the task from a machine learning (ML) perspective: we learn a parametric distribution $\tilde{p}_{\text{trn.}}(\symbolvar{y}_{\text{trn.}} | \symbolvar{\symbolvec{x}}_{\text{trn.}}, \Theta)$ given a training set $\symbolset{D}_{\text{trn.}}=\{\symbolvec{x}_{\text{trn}}^{(i)}, y_\text{trn}^{(i)} \}_i$, where $\Theta$ is the learnable parameter set. We expect good performance of using $\tilde{p}_{\text{trn.}}$ to classify any new test sample ${\symbolvec{x}_{\text{test}}}$ from a test set $\symbolset{D}_{\text{test}}$ if $\tilde{p}_{\text{trn.}}$ well approximates the true distribution ${p}_{\text{trn.}}(\symbolvar{y}_{\text{trn.}} | \symbolvar{\symbolvec{x}}_{\text{trn.}})$ \emph{and} the training and testing data are independent and \emph{identically} distributed, or ${p}_{\text{trn.}}(\symbolvar{y}_{\text{trn.}} | \symbolvar{\symbolvec{x}}_{\text{trn.}})\approx{p}_{\text{test}}(\symbolvar{y}_{\text{test}} | \symbolvar{\symbolvec{x}}_{\text{test}})$. 

\begin{figure}
    \centering
    \includegraphics[width=\linewidth, trim=10 379 10 83, clip]{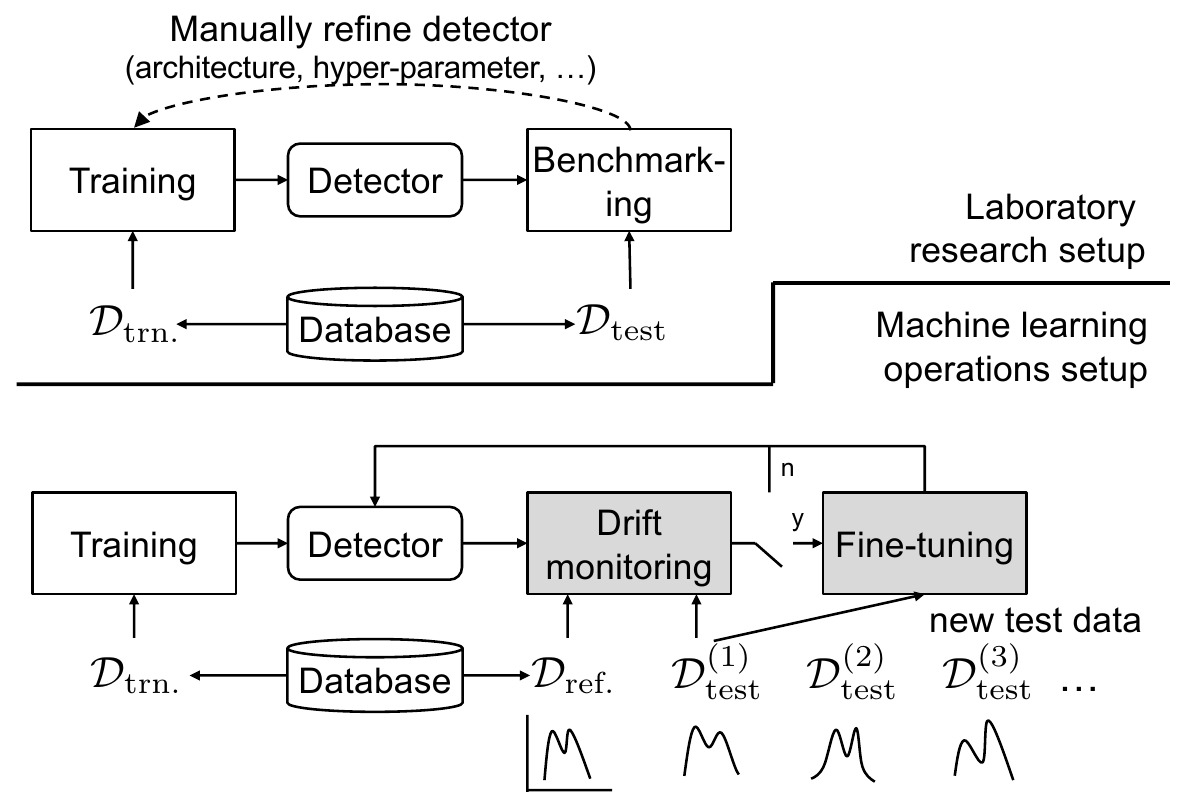}
    \vspace{-5mm}
    \caption{Illustration of speech deepfake detection in laboratory and MLOps setups. The highlighted blocks are investigated in this paper.}
    \label{fig:idea}
    \vspace{-2mm}
\end{figure}

In a common laboratory research setup (top panel of Fig.~\ref{fig:idea}), we are given a pair of $\symbolset{D}_{\text{trn}}$ and $\symbolset{D}_{\text{test}}$ from benchmarking databases~\cite{yiADD2022, todiscoASVspoof2019Future2019, challenge_safe2025},  
where $\symbolset{D}_{\text{test}}$ is usually designed to be slightly different from $\symbolset{D}_{\text{trn}}$. 
We then iterate over the loop of model training, testing, detector re-designing, and hyper-parameter tuning so that the learned $\tilde{p}_{\text{trn.}}$ better approximates ${p}_{\text{trn.}}$ and generalizes to the mismatched ${p}_{\text{test}}$ of $\symbolset{D}_{\text{test}}$. 
This setup accelerates the research iterations, and many effective solutions have been found, e.g., end-to-end deep neural network (DNN) classifiers~\cite{jung2022aasist} or their combination with self-supervised learning (SSL) speech feature extractors~\cite{takAutomaticSpeakerVerification2022}.

However, when deploying a deepfake detection application from the perspective of ML-operations (MLOps), we cannot assume that the detector trained on a training set created five years ago will perform well against the latest deepfakes, regardless of how well the detector `generalizes' to the test set paired with the training set in the laboratory research setup. More specifically, in the MLOps setup, we are facing a sequence of $(\symbolset{D}_{\text{test}}^{(1)}, \symbolset{D}_{\text{test}}^{(2)},\cdots)$ that may incrementally include deepfakes built upon newly proposed text-to-speech (TTS) or voice conversion (VC) algorithms. These new test sets may gradually become different, or \emph{drift away}, from the seen data. Monitoring the drift is hence helpful to signify the new and unseen incoming test data. When the drift is signified, we need to update the detector so that the drift can be reduced and the detection error rates will not be severely degraded. 

In the context of MLOps, we investigate the data drift caused by new deepfake attacks and focus on two research questions:
\begin{itemize}
\item[RQ1:] Whether drift caused by new TTS attacks can be monitored.
\item[RQ2:] Whether the drift can be reduced by fine-tuning the detector using similarly drifted data when it is available. 
\end{itemize} 
For RQ1, we monitor the drift by measuring the distance between the feature distributions of the incoming test data and a reference data (i.e., a development set). We conducted experiments using features extracted from SSL or non-SSL-based speech detectors on multiple datasets, including the constantly evolving MLAAD dataset~\cite{mullerMLAAD2024}. The results demonstrated that new TTS attacks indeed caused higher drift values than earlier ones. For RQ2, we demonstrated on the MLAAD dataset that the drift on the test set can be reduced if the detectors are fine-tuned using data synthesized from the new TTS attacks, and the degree of reduction is affected by the amount of fine-tuning data. Furthermore, fine-tuning reduces the detection error rates.

As far as we are aware of, this paper is the first to quantify the data drift of speech deepfake detection for MLOps. Other contributions include the comparison of multiple distance metrics over multiple datasets and detectors and the experiment of fine-tuning that demonstrates the reduction of drift and detection errors.

\section{Methods}
\label{sec:method}
\subsection{Definition of the drift}
 
Given
${p}(\symbolvar{y} | \symbolvar{\symbolvec{x}})= {p}( \symbolvar{\symbolvec{x}}|\symbolvar{y})\cdot{p}( \symbolvar{y}) / {p}( \symbolvar{\symbolvec{x}})$,\footnote{To avoid the notation clutter, we drop the suffix in $p_\text{test}$ and use $p$ to denote the distribution unless otherwise stated.} 
the change of ${p}(\symbolvar{\symbolvec{x}}|\symbolvar{y})$ is referred to as \emph{concept drift} in the previous artificial intelligence era~\cite{schlimmerIncremental1986}, wherein the word \emph{concept} refers to association between the data $\symbolvec{x}$ and its label $y$.  
For detecting speech deepfakes, we focus on the concept drift (or simply drift) of 
the fake data's conditional distribution 
${p}( \symbolvar{\symbolvec{x}}|\symbolvar{y} = \text{fake})$. The drift may happen when the speech data $\symbolvec{x}$ is generated using new TTS and VC systems.  

\subsection{Measuring the drift}
\label{sec:method:drift}
Inspired by existing studies~\cite{widmerLearning1996,zliobaiteOverview2016}, we define the drift of the fake data as a distance $D_{\text{t}-\text{r}}$ between ${p}( \symbolvar{\symbolvec{x}}|\symbolvar{y} = \text{fake})$ of a test set and the distribution ${p}_{\text{ref.}}( \symbolvar{\symbolvec{x}}_{\text{ref.}}|\symbolvar{y}_{\text{ref.}} = \text{fake})$ of a reference dataset $\symbolset{D}_{\text{ref}}$. This reference data can be $\symbolset{D}_{\text{trn.}}$, an early version of $\symbolset{D}_{\text{test}}$, or a dedicated development set (in the case of this paper).

Directly estimating $\tilde{p}(\symbolvar{\symbolvec{x}} | \symbolvar{y}, \Theta)$ is challenging because the waveform data is high-dimensional and varied in length. This is addressed with two approximations. First, we extract a fixed-dimensional embedding vector $\symbolvec{a}\in\symboldomain{R}^M$ from the waveform $\symbolvec{a}=\mathcal{H}_{\Theta}(\symbolvec{x})$, hoping that $\tilde{p}(\symbolvar{a} |\symbolvar{y})$ captures all the information in $\tilde{p}(\symbolvar{x} |\symbolvar{y})$. 
The embedding extractor $\mathcal{H}_\Theta$ is part of the detector, for example, the SSL-model front end~\cite{takAutomaticSpeakerVerification2022} plus a global average pooling layer. 
Although the embedding vector is more compact than a waveform (i.e., $M\ll T$), it is still challenging to accurately model $\tilde{p}(\symbolvar{a} |\symbolvar{y})$ for an $M$ larger than a few hundred. 

Hence, the second approximation is to model each dimension of $\symbolvec{a}$ independently. 
For the $i$-th dimension ${a}_i$, $i\in[1, M]$, we estimate the discrete probability mass function (PMF) $f$ (i.e., a histogram of ${a}_i$) or its cumulative distribution function (CDF) $F$. The same procedure is applied to estimate the PMF or CDF for the reference data ${a}_{\text{ref},i}$.   
Then, for example using the PMFs, we compute the drift distance between the test and reference data as $D_{\text{t}-\text{r}}\approx \sum_{i=1}^M D\Big({f}( {i}), {f}_{\text{ref.}}({i})\Big)$,
where $D$ is a distance function of two univariate PMFs, and ${f}({i})$ is the probability of taking the $i$-th value in the PMF support (i.e., the $i$-th histogram bin). 

We compare three representative distance functions~\cite{zliobaiteOverview2016} listed in Table~\ref{tab:metrics}. Note that the Kolmogoro-Smirnov (K-S) test was defined from the perspective of a statistical test, even though its implementation is similar to the Wasserstein-1 distance. 

\input{table_metrics}

\subsection{Fine-tuning the detector}
In MLOps, if a test dataset is judged to be drifting away from the reference, the model can be updated using the test data after the labels are annotated. 
The annotated data can be merged with the original training data, or only useful data is selected to fine-tune the model via active learning~\cite{settlesActiveLearningLiterature2009}. Even more straightforwardly, new data for fine-tuning the detector can be randomly sampled. 

This paper conducts fine-tuning using randomly sampled new data because the implementation cost is low, and its performance is only slightly inferior to more complicated active learning algorithms~\cite{wangInvestigatingActivelearningbasedTraining2023}. Furthermore, for the experiment in this paper, we make sure that the randomly sampled utterances for fine-tuning are different from the test set utterances, even though they are from the same set of TTS and VC attacks.
Note again that the MLOps setup is different from the laboratory setup, wherein the attacks in the fine-tuning and test sets are different.

\section{Experiments}
\label{sec:exp}

\subsection{Design of experiments}
The first experiment addresses \emph{whether the drift can be monitored} (RQ1). The sequence of test datasets $(\symbolset{D}_{\text{test}}^{(1)}, \symbolset{D}_{\text{test}}^{(2)},\cdots)$ is created using a carefully curated single-speaker dataset or the English portion of the multi-speaker MLAAD database~\cite{mullerMLAAD2024}. 
We exhaustively combine the three distance metrics in Table~\ref{tab:metrics} with three pre-trained detectors, including two SSL-based detectors in different sizes~\cite{gePosttraining2025} and a non-SSL detector called AASIST~\cite{jung2022aasist}. We then measure the drift on the test sets using the distance metrics and the features extracted by the detectors. 
The experiment design also allows us to investigate the impact of the distance metric, the type of detector, and the dataset itself when monitoring the drift.

We conduct the second experiment on fine-tuning the detector (RQ2) using the MLAAD database. Instead of creating an actual MLOps loop and directly fine-tuning and evaluating on the same $\symbolset{D}_{\text{test}}^{(m)}$, we use sufficiently varied experiment settings to investigate the performance 1) when amount of fine-tuning data varies and 2) when the attacks in the fine-tuning and test data mismatch. First, for each $\symbolset{D}_{\text{test}}^{(m)}$, multiple fine-tuning sets $\{\symbolset{D}_{\text{ft.}}^{(m,k)}\}_k$ are created, where each $\symbolset{D}_{\text{ft.}}^{(m,k)}$ covers the same set of attacks as $\symbolset{D}_{\text{test}}^{(m)}$ but contains a different number of utterances disjoint\footnote{The MLAAD database does not provide speaker labels. Hence we did not make the speakers in the fine-tuning and test sets disjoint.} from those in $\symbolset{D}_{\text{test}}^{(m)}$. 
For each condition $(m, k)$, we fine-tune the detector using $\symbolset{D}_{\text{ft.}}^{(m,k)}$ and measure the drift value on $\symbolset{D}_{\text{test}}^{(m)}$. This setting is expected to reveal the impact of the amount of fine-tuning data. We also measure the drift on other test sets $\symbolset{D}_{\text{test}}^{(n\neq m)}$, which demonstrates the performance when the fine-tuning and test data have different attacks.

\subsection{Datasets} 
For the single-speaker dataset used in the first experiment, we collect a LJSpeech-TTS dataset consisting of 1,881 synthetic utterances (3.3 hours) from 12 TTS systems created in the last eight years: the ESPNet version~\cite{hayashi2020espnet} of Tacotron v1~\cite{wangTacotron2017} and v2~\cite{shen2018natural}, Transformer-TTS~\cite{li2019neural}, FastSpeech series~\cite{ren2019fastspeech,ren2020fastspeech}, VITS~\cite{kim2021conditional}, and other latest TTS systems. 
A $\symbolset{D}_{\text{test}}^{(m)}$ is created for each of the 12 TTS systems. All the TTS systems were built using the female voice in the LJSpeech database~\cite{ljspeech17}. The synthetic utterances are downloaded from the TTS systems' demonstration pages.
As for the second dataset, we use the English subset of the multi-speaker MLAAD dataset version 7.0, which covers 54 TTS systems built in the past eight years. Each $\symbolset{D}_{\text{test}}^{(m)}, \forall m\in[1, 54]$, contains five hours' data randomly sampled from the corresponding TTS attack in MLAAD.

Experiment 2 is conducted on the MLAAD dataset. The test sets are the same as those of experiment 1, and we create the fine-tuning datasets using the remaining utterances in the MLAAD dataset. 
Ideally, we need to create $\symbolset{D}_{\text{ft.}}^{(m, k)}$, where $m\in[1, 54]$. To reduce the experimental cost, we group the TTS systems on the basis of their corresponding MLAAD database version ID from \symboltag{v2} to \symboltag{v7}\footnote{New TTS systems were added when MLAAD was version upgraded, except the case from version 1.0 to 2.0. Hence, we use tags from \symboltag{v2} to \symboltag{v7}} and create the version-wise fine-tuning sets $\symbolset{D}_{\text{ft.}}^{(m', k)}$, where $m'\in\{\symboltag{v2}, \cdots, \symboltag{v7}\}$. 
For each $m'$, for example $m'=\symboltag{v7}$, we randomly sample 0.5, 2, 4, or 8 hours' data from the TTS attacks with label \symboltag{v7}, which leads to four fine-tuning sets $\{\symbolset{D}_{\text{ft.}}^{(\symboltag{v7}, 0.5)}, \cdots \symbolset{D}_{\text{ft.}}^{(\symboltag{v7}, 8)}\}$ for $m'=\symboltag{v7}$. Furthermore, for supervised fine-tuning of the detector, the same amount of real human speech data is sampled from the real version of MLAAD (i.e., the M-AILABS dataset~\cite{m-ailabsMAILABS}) and added to the corresponding fine-tuning set. 
The procedure is also used to create the fine-tuning sets for versions from \symboltag{v2} to \symboltag{v6}.

The reference data is the ASVspoof 2019 development set for both experiments. 
When the real data is needed to compute the detection equal error rate (EER) on each $\mathcal{D}_{\text{test}}$, we sample the same amount of real data from the LJspeech or M-AILABS dataset as the TTS data in the test sets. All the data has a sampling rate of 16 kHz.

\subsection{Detector configurations and recipes}
\label{sec:exp:models}
The three detectors are configured as below: 
\begin{itemize}
\item \mAASIST{}~\cite{jung2022aasist}: an end-to-end detector that combines a sinc-filterbank and a graph-attention DNN. It uses the official implementation with around 300k trainable parameters. 
\item \mWVS{}: an SSL-based detector with a small wav2vec 2.0 module~\cite{NEURIPS2020_92d1e1eb} as the front end and a shallow back end using a global average pooling layer and a linear output layer. 
This detector is implemented using the AntiDeepfake toolkit~\cite{gePosttraining2025} and has around 95 million parameters. 
\item \mXLSR{}: similar to \mWVS{} but with a large SSL module called XLS-R ~\cite{babu2021xls}. The detector has around 2 billion parameters. 
\end{itemize} 
We use \mXLSR{} because of its top-tier performance on various datasets~\cite{gePosttraining2025,takAutomaticSpeakerVerification2022}. The smaller \mWVS{} is less powerful but useful for memory-constrained applications. We also use \mAASIST{} since it is a top-performing detector among those without SSL-based modules.

The embedding vector ($\symbolvec{a}$ in Sec.~\ref{sec:method:drift}) is extracted from the layer before the last linear layer of each detector. The number of dimensions is 160, 768, and 1,920 for the three detectors, respectively. 
We initialize \mAASIST{} using the pre-trained checkpoint from the official repository, which was trained on the ASVspoof 2019 LA training set~\cite{todiscoASVspoof2019Future2019}. The checkpoints of the SSL-based detectors were trained using the AntiDeepfake recipe on the ASVspoof 2019 and ASVspoof 5 training sets~\cite{wang2024asvspoof5}. 

For experiment 2, we fine-tune the pre-trained detectors for five epochs using an AdamW optimizer ($\beta_1=0.9, \beta_2=0.999, \epsilon=10^{-8}$, weight decay of 0.01). The learning rates of \mAASIST{}, \mWVS{}, and \mXLSR{} are $1e-4$, $1e-6$, and $1e-7$, respectively. 

\subsection{Experiment 1: Can we monitor data drift?}
\label{sec:exp:result1}

\input{figure_e1}

Figure~\ref{fig:e1} presents the drift values measured using the three detectors on the LJSpeech-TTS dataset. The TTS systems are sorted on the horizontal axis on the basis of their publication date, assuming that the ordering generally correlates with the technical progress of the TTS systems. The drift values measured using the pre-trained detectors on the MLAAD test set are plotted using the bold grey profile in Fig.~\ref{fig:e2}. Note that we computed the drift for each individual TTS attack but do not label all the TTS systems due to limited space. 

First, as the figures indicate, \textbf{the drift is monitored in the sense that the latest TTS systems tend to cause higher drift values than the early ones}. On the LJSpeech-TTS dataset, the JETS~\cite{limJETS2022a}, NaturalSpeech~\cite{tanNaturalspeech2024}, and StyleTTS2~\cite{liStyleTTS} tend to yield higher drift values than the `earlier' systems based on Tacotron, Transformer, FastSpeech, and VITS on the three detectors. A potential reason is that the new systems are more advanced: JETS jointly optimizes FastSpeech and its waveform generator; StyleTTS2 introduces SSL-based adversarial training to FastSpeech; NaturalSpeech improved the prior and posterior used in VITS.
On the MLAAD dataset, we also observe higher drift values on TTS data  from \symboltag{v6} and \symboltag{v7}, which are more advanced than those in the earlier versions of MLAAD.

Figure~\ref{fig:e1} also illustrates the EER per attack. The observation is that the drift values show similar patterns across the three detectors, even though the EERs are dramatically different. When using \mXLSR{} although the EERs are 0\% on all the TTS attacks, the drift values of the latest TTS are still higher than the earlier systems. 
This is not unreasonable because what EER and drift values measure are different --- 
The EER is about discriminating the TTS-generated data from the real data (i.e., LJSpeech) while the drift is about how far away the TTS data is from the reference data. 

Finally, the three distance functions produce similar results. The pair-wise correlation of the drift values of different metrics is higher than 0.8. The choice of distance function may not be critical, and we only present the results using the Wasserstein-1 distance from now.

\input{figure_e2}

\subsection{Experiment 2: Can we reduce the drift via fine-tuning?}
\label{sec:exp:result2}
We fine-tune the detectors using each of the fine-tuning sets $\{\mathcal{D}_{\text{ft.}}^{(\symboltag{v2}, 1)}, \cdots\mathcal{D}_{\text{ft.}}^{(\symboltag{v7}, 4)}\}$ and compute the drift on all the test sets. 
Due to the limited space, we only plot in Fig.~\ref{fig:e2} the drift values for \mXLSR{} using  
$\mathcal{D}_{\text{ft.}}^{(\symboltag{m'}, k)}$, $\forall m'\in\{\symboltag{v7}, \symboltag{v6}, \symboltag{v2}\}$, and $\forall k\in\{0.5, 2, 4, 8\}$.
Other results are in the appendix.\footnote{Please check link for appendix and code repository.}

The first message is that \textbf{fine-tuning the detector with new TTS data is likely to reduce the drift caused by test data from the TTS systems.} Specifically, 
Fig.~\ref{fig:e2}(a) shows that, when fine-tuning the detector using 8 hours' data from \symboltag{v7}, the high drift values on the TTS attacks in \symboltag{v7} decrease. Furthermore, the decrease is more obvious when using more fine-tuning data. This trend can also be observed from the results using fine-tuning sets of \symboltag{v6} (Fig.~\ref{fig:e2}(b)). This is not surprising since the detector is fine-tuned using the data from the same sets of TTS systems.

What is interesting is that, as shown on the right hand side of Fig.~\ref{fig:e2}(b), the detector fine-tuned with data of \symboltag{v6} also reduce the drift on the \symboltag{v7} test data, even though the TTS attacks in \symboltag{v6} are different from those in \symboltag{v7}. A potential reason could be that the TTS attacks in the two sets use similar techniques such as diffusion. 

However, as Fig.~\ref{fig:e2}(c) shows, \textbf{fine-tuning using the `old' data of \symboltag{v2} does not reduce the drift caused by newer TTS attacks}; neither does it reduce the already-low drift values on the \symboltag{v2} test data. Although not presented in this paper, using the fine-tuning data from \symboltag{v3}, \symboltag{v4}, and \symboltag{v5} leads to similar results to what Fig.~\ref{fig:e2}(c) shows.

In addition to the drift values, we are curious about how the EERs change across the fine-tuning conditions. 
To save space, we pool the EERs on the TTS attacks with the same version ID. As the bottom two rows of Table~\ref{tab:eer} show, the EERs on \symboltag{v7} and \symboltag{v6} test set are reduced when the detector is fine-tuned with a fine-tuning set from either \symboltag{v7} or \symboltag{v6}. The decrease is more significant when using more data. Using 8 hours' fine-tuning data from \symboltag{v2} also reduces the EER on \symboltag{v6} (5.40\% $\rightarrow$ 2.65\%), but the decrease is less than that when using the 8 hours' fine-tuning set from \symboltag{v6} (5.18\% $\rightarrow$ 1.19\% ) or \symboltag{v7} (5.04\% $\rightarrow$ 0.96\% ). The same trend is observed on the \symboltag{v7} test set.

The results of the two experiments suggest that the drift caused by new TTS systems can be measured and observed. To reduce the drift and potentially improve the detection performance, fine-tuning the detector with data from a similar source is promising.

\input{table_eer_ver4}

\section{Conclusions}
We investigated the monitoring of data drift for detecting speech deepfakes in machine learning operations (MLOps). Experiments on multiple datasets suggest that the drift can be monitored using the distance between the feature distributions of the test data and reference data. Furthermore, the drift values caused by newer text-to-speech (TTS) attacks tend to be larger than those caused by the relatively old-fashioned attacks. 
Simulating the development flow in MLOps, we also found that fine-tuning the detector using new TTS data can help it be more robust to test data from similar attacks.

On the basis of drift detection, we plan to investigate more efficient fine-tuning methods (e.g., LoRA~\cite{hu_lora_2022}) rather than fine-tuning the whole detector. This may reduce the fine-tuning data from 4 or 8 hours to a more reasonable duration. Another topic is to extend the evaluation metric on static test sets so that the detection performance can be fairly compared across the evolving test sets.

\vfill\pagebreak

\pretolerance=1000

\bibliographystyle{IEEEbib}
\bibliography{refs}
\input{app}
\end{document}

%% file: table_metrics.tex
\begin{table}[t!]
    \centering
    \caption{Investigated distance metrics $D_{p-q}$ for discrete univariate PMFs ($f_p$, $f_q$) or CDFs ($F_p$, $F_q$). The ordered support of the discrete PMF or CDF is written as $\{s_1, \cdots, s_N\} \in\symboldomain{R}$.}
        \vspace{-3mm}
    \resizebox{0.49\textwidth}{!}{
    \begin{tabular}{rl}
    \toprule
    Name & Definition of $D_{p-q}$ \\
    \midrule
     Wasserstein-1 distance & $ \sum_{n=2}^{N} \Big|F_p(s_n) - F_q(s_n) \Big|(s_n - s_{n-1})$ \\
     Kolmogoro-Smirnov (K-S) test & $\max_n \Big|F_p(s_n) - F_q(s_n) \Big|$ \\
     Kullback-Leibler diver. (KLD) & $\sum_{n=i}^{N} f_p \log\Big(f_p(s_n) / f_q (s_n)\Big)$ \\
     \bottomrule
    \end{tabular}
    }
    \label{tab:metrics}
    
\end{table}

%% file: figure_e1.tex
\begin{figure}[t!]
    \centering
    \begin{subfigure}[t]{0.45\textwidth}
        \vskip 0pt
        \centering
        \includegraphics[width=\textwidth, trim=0 55 0 0, clip]{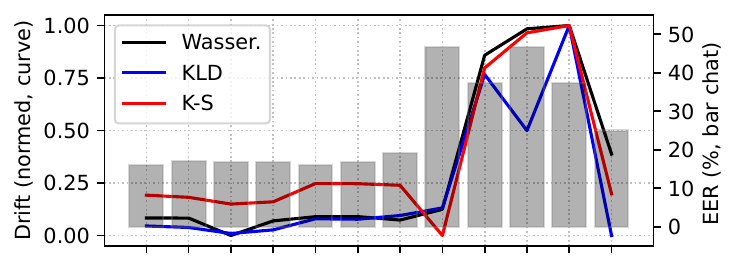}
        \caption{\mAASIST{}}
        \label{fig:e1:11}
    \end{subfigure}
    \hfill
    \begin{subfigure}[t]{0.45\textwidth}
        \vskip 0pt
        \centering
        \includegraphics[width=\textwidth, trim=0 55 0 0, clip]{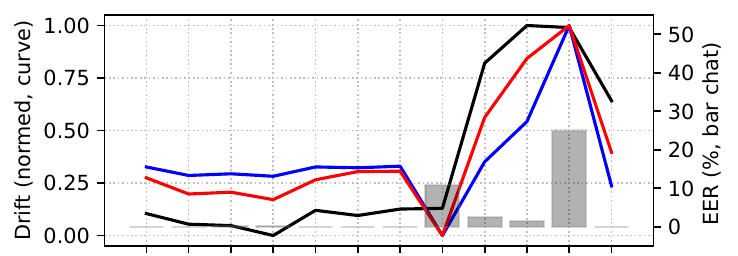}
        \caption{\mWVS{}}
        \label{fig:e1:21}
    \end{subfigure}
    \hfill
    \begin{subfigure}[t]{0.45\textwidth}
        \vskip 0pt
        \centering
        \includegraphics[width=\textwidth, trim=0 10 0 0, clip]{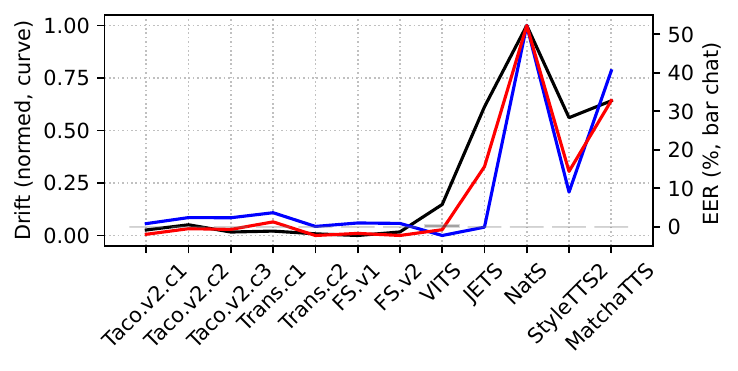}
        \caption{\mXLSR{}}
        \label{fig:e1:31}
    \end{subfigure}
    \vspace{-2mm}
    \caption{Normalized drift values (curves) computed on the TTS data from the LJSpeech-TTS dataset, using the three distance functions (Sec.~\ref{sec:method}) and the three detectors (Sec.~\ref{sec:exp:models}). Each curve is min-max normalized in order to fit the same numeric range for visualization. EERs (bar chats) are computed using the TTS data and the corresponding real data. The TTS systems are chronologically ordered from left to right based on the paper's publication date.}
    \label{fig:e1}
    \vspace{-2mm}
\end{figure}

%% file: figure_e2.tex
\begin{figure}[t!]
    \centering
    \begin{subfigure}[t]{0.45\textwidth}
        \vskip 0pt
        \centering
        \includegraphics[width=\textwidth]{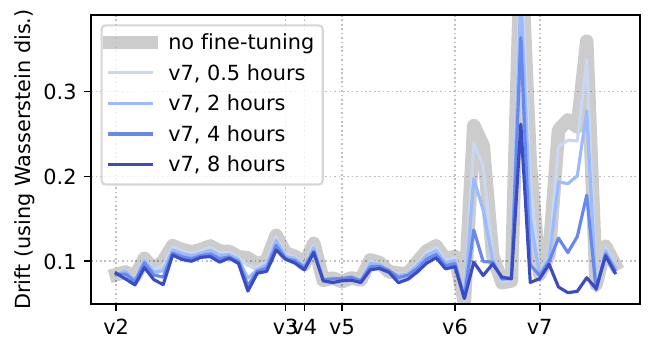}
        \vspace{-5mm}
        \caption{Using \symboltag{v7} fine-tuning sets $\{\symbolset{D}_{\text{ft.}}^{(\symboltag{v7}, 0.5)}, \symbolset{D}_{\text{ft.}}^{(\symboltag{v7}, 2)}, \symbolset{D}_{\text{ft.}}^{(\symboltag{v7}, 4)}, \symbolset{D}_{\text{ft.}}^{(\symboltag{v7}, 8)}\}$}
        \label{fig:e2:1}
    \end{subfigure}
    \hfill
    \begin{subfigure}[t]{0.45\textwidth}
        \vskip 0pt
        \centering
        \includegraphics[width=\textwidth]{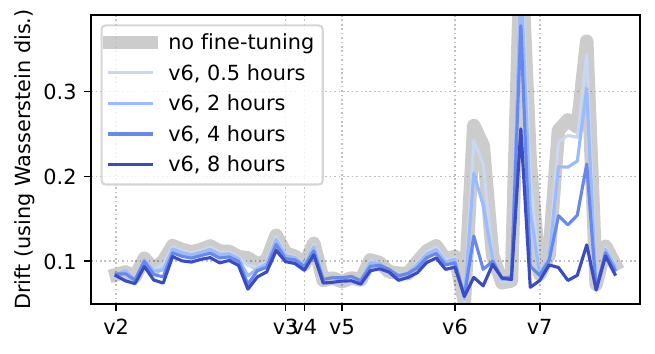}
        \vspace{-5mm}
        \caption{Using \symboltag{v6} fine-tuning sets $\{\symbolset{D}_{\text{ft.}}^{(\symboltag{v6}, 0.5)}, \symbolset{D}_{\text{ft.}}^{(\symboltag{v6}, 2)}, \symbolset{D}_{\text{ft.}}^{(\symboltag{v6}, 4)} ,\symbolset{D}_{\text{ft.}}^{(\symboltag{v6}, 8)}\}$}
        \label{fig:e2:2}
    \end{subfigure}
    \hfill
    \begin{subfigure}[t]{0.45\textwidth}
        \vskip 0pt
        \centering
        \includegraphics[width=\textwidth]{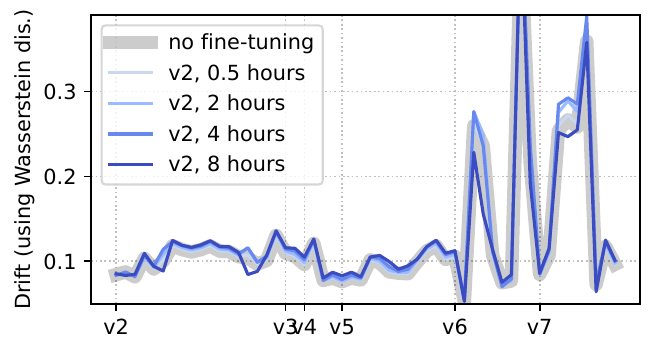}
        \vspace{-5mm}
        \caption{Using \symboltag{v2} fine-tuning sets $\{\symbolset{D}_{\text{ft.}}^{(\symboltag{v2}, 0.5)}, \symbolset{D}_{\text{ft.}}^{(\symboltag{v2}, 2)}, \symbolset{D}_{\text{ft.}}^{(\symboltag{v2}, 4)},\symbolset{D}_{\text{ft.}}^{(\symboltag{v2}, 8)}\}$}
        \label{fig:e2:3}
    \end{subfigure}
    \vspace{-2mm}
    \caption{Drift values measured on MLAAD TTS systems using Wasserstein-1 distance and pre-trained (grey profile) or fine-tuned \mXLSR{} (colored blue profiles). 
    The 54 TTS systems are sorted based on their MLAAD version IDs and the paper's publication date. Names of TTS systems are not shown.
    }
    \label{fig:e2}
    \vspace{-2mm}
\end{figure}

%% file: table_eer_ver4.tex
\begin{table}[!t]
\centering
    \caption{EERs on \symboltag{v2}, \symboltag{v6}, and \symboltag{v7} test sets (rows) when the detector \mXLSR{} was fine-tuned using different fine-tuning sets (columns). The condition with no fine-tuning is listed in the 2nd column from left. 
    A darker color indicates a higher EER value. }
    \label{tab:eer}
     \vspace{-2mm}
\resizebox{0.49\textwidth}{!}{
\setlength{\tabcolsep}{1pt}
\begin{tabular}{c|c|cccc|cccc|cccc}     
\toprule
test& no &\multicolumn{4}{c}{\symboltag{v2} fine-tuning set} & \multicolumn{4}{c}{\symboltag{v6} fine-tuning set}& \multicolumn{4}{c}{\symboltag{v7} fine-tuning set} \\ \cmidrule(lr){3-6} \cmidrule(lr){7-10}\cmidrule(lr){11-14}
set   &  ft.  & 0.5h & 2.0h & 4.0h & 8.0h & 0.5h & 2.0h & 4.0h & 8.0h & 0.5h & 2.0h & 4.0h & 8.0h\\ 
\midrule
\symboltag{v2} & \cellcolor[rgb]{0.99, 0.99, 0.99} 0.40 & \cellcolor[rgb]{0.99, 0.99, 0.99} 0.39 & \cellcolor[rgb]{0.99, 0.99, 0.99} 0.35 & \cellcolor[rgb]{0.99, 0.99, 0.99} 0.26 & \cellcolor[rgb]{1.00, 1.00, 1.00} 0.05  & \cellcolor[rgb]{0.99, 0.99, 0.99} 0.46 & \cellcolor[rgb]{0.99, 0.99, 0.99} 0.38 & \cellcolor[rgb]{0.99, 0.99, 0.99} 0.28 & \cellcolor[rgb]{0.99, 0.99, 0.99} 0.23 & \cellcolor[rgb]{0.99, 0.99, 0.99} 0.46 & \cellcolor[rgb]{0.99, 0.99, 0.99} 0.37 & \cellcolor[rgb]{0.99, 0.99, 0.99} 0.24 & \cellcolor[rgb]{1.00, 1.00, 1.00} 0.05\\ 
\symboltag{v6} & \cellcolor[rgb]{0.70, 0.70, 0.70} 5.40 & \cellcolor[rgb]{0.71, 0.71, 0.71} 5.36 & \cellcolor[rgb]{0.72, 0.72, 0.72} 5.22 & \cellcolor[rgb]{0.81, 0.81, 0.81} 4.02 & \cellcolor[rgb]{0.89, 0.89, 0.89} 2.65  & \cellcolor[rgb]{0.72, 0.72, 0.72} 5.18 & \cellcolor[rgb]{0.76, 0.76, 0.76} 4.63 & \cellcolor[rgb]{0.87, 0.87, 0.87} 2.93 & \cellcolor[rgb]{0.96, 0.96, 0.96} 1.19 & \cellcolor[rgb]{0.73, 0.73, 0.73} 5.04 & \cellcolor[rgb]{0.78, 0.78, 0.78} 4.42 & \cellcolor[rgb]{0.90, 0.90, 0.90} 2.44 & \cellcolor[rgb]{0.97, 0.97, 0.97} 0.96\\ 
\symboltag{v7} & \cellcolor[rgb]{0.61, 0.61, 0.61} 6.42 & \cellcolor[rgb]{0.61, 0.61, 0.61} 6.38 & \cellcolor[rgb]{0.61, 0.61, 0.61} 6.39 & \cellcolor[rgb]{0.65, 0.65, 0.65} 5.92 & \cellcolor[rgb]{0.81, 0.81, 0.81} 3.99 & \cellcolor[rgb]{0.61, 0.61, 0.61} 6.37 & \cellcolor[rgb]{0.68, 0.68, 0.68} 5.64 & \cellcolor[rgb]{0.86, 0.86, 0.86} 3.22 & \cellcolor[rgb]{0.95, 0.95, 0.95} 1.52 & \cellcolor[rgb]{0.61, 0.61, 0.61} 6.35 & \cellcolor[rgb]{0.71, 0.71, 0.71} 5.36 & \cellcolor[rgb]{0.91, 0.91, 0.91} 2.23 & \cellcolor[rgb]{0.98, 0.98, 0.98} 0.57\\ 
\bottomrule
\end{tabular}
}
\vspace{-5mm}
\end{table}

%% file: app.tex
\newpage
\clearpage
\appendix
\section{Appendix}

\subsection{Additional result of using pre-trained models on MLAAD}

\begin{figure}[h!]
    \centering
    \begin{subfigure}[t]{0.45\textwidth}
        \vskip 0pt
        \centering
        \includegraphics[width=\textwidth, trim=0 20 0 0, clip]{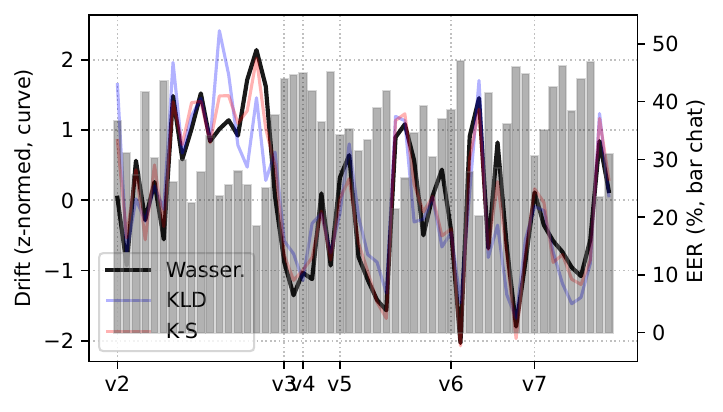}
        \caption{\mAASIST{}}
        \label{fig:e1:11}
    \end{subfigure}
    \hfill
    \begin{subfigure}[t]{0.45\textwidth}
        \vskip 0pt
        \centering
        \includegraphics[width=\textwidth, trim=0 20 0 0, clip]{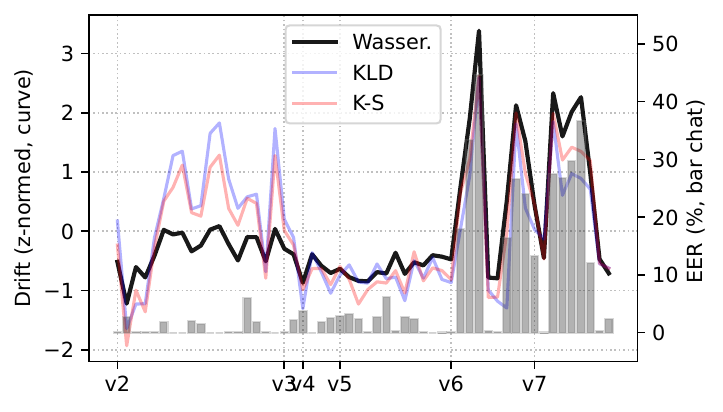}
        \caption{\mWVS{}}
        \label{fig:e1:21}
    \end{subfigure}
    \hfill
    \begin{subfigure}[t]{0.45\textwidth}
        \vskip 0pt
        \centering
        \includegraphics[width=\textwidth, trim=0 0 0 0, clip]{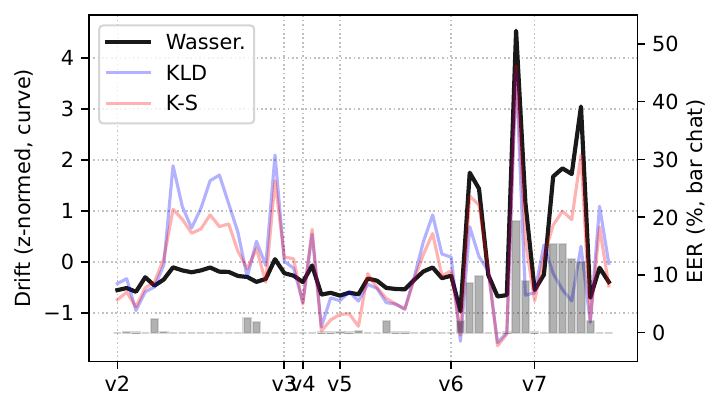}
        \caption{\mXLSR{}}
        \label{fig:e1:31}
    \end{subfigure}
    \vspace{-2mm}
    \caption{Normalized drift values computed on the MLAAD test set, using the three distance functions (Sec.~\ref{sec:method}) and the three detectors with no fine-tuning (Sec.~\ref{sec:exp:models}). Each curve is min-max normalized to fit the same numeric range for visualization. The TTS systems are ordered in the same way as in Fig.~\ref{fig:e2}. The drift value curve computed using the Wasserstein-1 distance (in black) is the normalized version of  the `no fine-tuning' curve (grey profile) in Fig.~\ref{fig:e2}.}
    \label{fig:app_e1}
    \vspace{-2mm}
\end{figure}

Similar to Fig.~\ref{fig:e1} on the toy TTS dataset, in Fig.~\ref{fig:app_e1}, we plot and compare the results of using three distance functions on the MLAAD test set. The three detectors are pre-trained without fine-tuning, the same as those in Fig.~\ref{fig:e1}.  

We list a few observations:

\begin{itemize}
\item The drift values from the two SSL models show similar patterns, but those from \mAASIST{} look random. The EERs of \mAASIST{} are around 20\% for all the attacks. On more varied test data, detectors with better capacity may be needed to compute useful drift values. 
\item Similar to the observation from Sec.~\ref{sec:exp:result1}, the three distance functions show similar patterns across different versions of the MLAAD test set. However, the KLD and K-S seem to produce higher drift values on some of the attacks in MLAAD \symboltag{v2}. The reasons remain unknown, but those attacks are all pre-trained TTS models from the ESPNet and applied to the speakers in the MLAAD database.
\end{itemize}

\subsection{Additional results of fine-tuning \mXLSR{} on MLAAD}

Following Fig.~\ref{fig:e2}, we present the results using the fine-tuning data of \symboltag{v3}, \symboltag{v4}, and \symboltag{v5} on \mXLSR.  Because \symboltag{v3} and \symboltag{v4} of MLAAD have few data, we only have $\{\symbolset{D}_{\text{ft.}}^{(\symboltag{v4}, 0.5)}, \symbolset{D}_{\text{ft.}}^{(\symboltag{v4}, 2)}, \symbolset{D}_{\text{ft.}}^{(\symboltag{v4}, 4)}\}$ for \symboltag{v4} and $\{\symbolset{D}_{\text{ft.}}^{(\symboltag{v3}, 0.5)}, \symbolset{D}_{\text{ft.}}^{(\symboltag{v3}, 2)}\}$ for \symboltag{v3}. 
Fine-tuning data from \symboltag{v3}, \symboltag{v4}, and \symboltag{v5} of MLAAD does not reduce the drift values.
\begin{figure}[t!]
    \centering
    \begin{subfigure}[t]{0.45\textwidth}
        \vskip 0pt
        \centering
        \includegraphics[width=\textwidth]{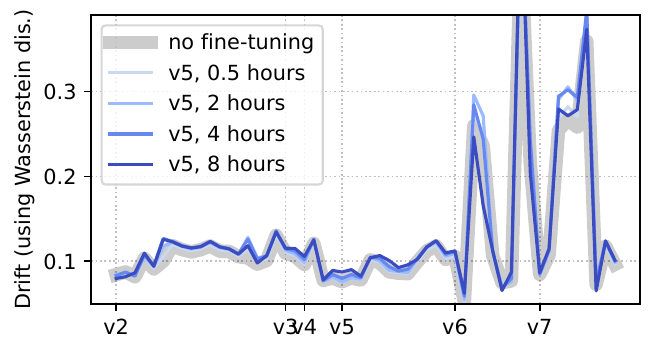}
        \vspace{-5mm}
        \caption{Using \symboltag{v5} fine-tuning sets $\{\symbolset{D}_{\text{ft.}}^{(\symboltag{v5}, 0.5)}, \symbolset{D}_{\text{ft.}}^{(\symboltag{v5}, 2)}, \symbolset{D}_{\text{ft.}}^{(\symboltag{v5}, 4)}, \symbolset{D}_{\text{ft.}}^{(\symboltag{v7}, 8)}\}$}
        \label{fig:e2:1}
    \end{subfigure}
    \hfill
    \begin{subfigure}[t]{0.45\textwidth}
        \vskip 0pt
        \centering
        \includegraphics[width=\textwidth]{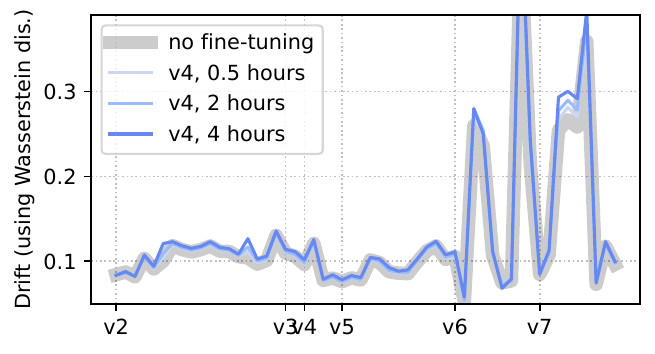}
        \vspace{-5mm}
        \caption{Using \symboltag{v4} fine-tuning sets $\{\symbolset{D}_{\text{ft.}}^{(\symboltag{v4}, 0.5)}, \symbolset{D}_{\text{ft.}}^{(\symboltag{v4}, 2)}, \symbolset{D}_{\text{ft.}}^{(\symboltag{v4}, 4)}\}$}
        \label{fig:e2:2}
    \end{subfigure}
    \hfill
    \begin{subfigure}[t]{0.45\textwidth}
        \vskip 0pt
        \centering
        \includegraphics[width=\textwidth]{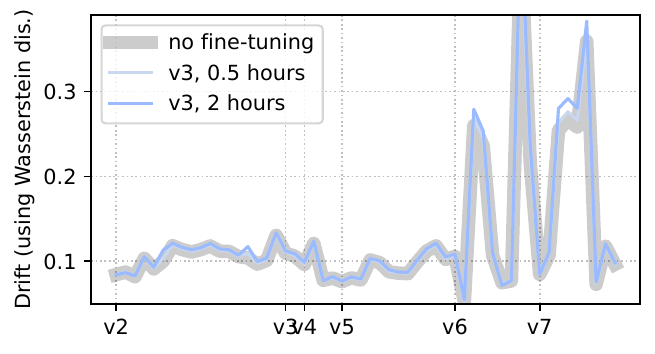}
        \vspace{-5mm}
        \caption{Using \symboltag{v3} fine-tuning sets $\{\symbolset{D}_{\text{ft.}}^{(\symboltag{v3}, 0.5)}, \symbolset{D}_{\text{ft.}}^{(\symboltag{v3}, 2)}\}$}
        \label{fig:e2:3}
    \end{subfigure}
    \vspace{-2mm}
    \caption{Drift values measured on MLAAD TTS systems using Wasserstein-1 distance and pre-trained (grey profile) or fine-tuned \mXLSR{} (blue profiles).  The 54 TTS systems are sorted on the basis of their MLAAD version IDs and the paper's publication date. Names of TTS systems are not shown. This figure is complementary to Fig.~\ref{fig:e2}.
    }
    \label{fig:e2}
    \vspace{-2mm}
\end{figure}

\begin{table*}[!t]
\centering
    \caption{EERs on versions of MLAAD test sets (rows) when the detector \mXLSR{} was fine-tuned using different fine-tuning sets (columns). The condition with no fine-tuning is listed in the 2nd column from the left.  Darker colors indicate a higher EER values. Note that \symboltag{v3} and \symboltag{v4} do not have all the fine-tuning conditions due to lack of data.}
    \label{tab:app:eer}
     \vspace{-2mm}
\resizebox{0.99\textwidth}{!}{
\setlength{\tabcolsep}{1pt}
\begin{tabular}{c|c|cccc|cccc|cccc|cccc|cccc|cccc}
\toprule
test& no &\multicolumn{4}{c}{\symboltag{v2} fine-tuning set} & \multicolumn{4}{c}{\symboltag{v3} fine-tuning set} & \multicolumn{4}{c}{\symboltag{v4} fine-tuning set} & \multicolumn{4}{c}{\symboltag{v5} fine-tuning set} &   \multicolumn{4}{c}{\symboltag{v6} fine-tuning set}& \multicolumn{4}{c}{\symboltag{v7} fine-tuning set} \\ \cmidrule(lr){3-6} \cmidrule(lr){7-10}\cmidrule(lr){11-14}\cmidrule(lr){15-18}\cmidrule(lr){19-22}\cmidrule(lr){23-26}
set   &  ft.  & 0.5h & 2.0h & 4.0h & 8.0h & 0.5h & 2.0h & \phantom{0.00} & \phantom{0.00} & 0.5h & 2.0h & 4.0h & \phantom{0.00} & 0.5h & 2.0h & 4.0h & 8.0h & 0.5h & 2.0h & 4.0h & 8.0h & 0.5h & 2.0h & 4.0h & 8.0h\\ 
\midrule
v2 & \cellcolor[rgb]{0.99, 0.99, 0.99} 0.40 & \cellcolor[rgb]{0.99, 0.99, 0.99} 0.39 & \cellcolor[rgb]{0.99, 0.99, 0.99} 0.35 & \cellcolor[rgb]{0.99, 0.99, 0.99} 0.26 & \cellcolor[rgb]{1.00, 1.00, 1.00} 0.05 & \cellcolor[rgb]{0.99, 0.99, 0.99} 0.37 & \cellcolor[rgb]{0.99, 0.99, 0.99} 0.28 &      &      & \cellcolor[rgb]{0.99, 0.99, 0.99} 0.37 & \cellcolor[rgb]{0.99, 0.99, 0.99} 0.35 & \cellcolor[rgb]{0.99, 0.99, 0.99} 0.26 &      & \cellcolor[rgb]{0.99, 0.99, 0.99} 0.37 & \cellcolor[rgb]{0.99, 0.99, 0.99} 0.28 & \cellcolor[rgb]{0.99, 0.99, 0.99} 0.27 & \cellcolor[rgb]{0.99, 0.99, 0.99} 0.21 & \cellcolor[rgb]{0.99, 0.99, 0.99} 0.46 & \cellcolor[rgb]{0.99, 0.99, 0.99} 0.38 & \cellcolor[rgb]{0.99, 0.99, 0.99} 0.28 & \cellcolor[rgb]{0.99, 0.99, 0.99} 0.23 & \cellcolor[rgb]{0.99, 0.99, 0.99} 0.46 & \cellcolor[rgb]{0.99, 0.99, 0.99} 0.37 & \cellcolor[rgb]{0.99, 0.99, 0.99} 0.24 & \cellcolor[rgb]{1.00, 1.00, 1.00} 0.05\\ 
v3 & \cellcolor[rgb]{1.00, 1.00, 1.00} 0.00 & \cellcolor[rgb]{1.00, 1.00, 1.00} 0.00 & \cellcolor[rgb]{1.00, 1.00, 1.00} 0.00 & \cellcolor[rgb]{1.00, 1.00, 1.00} 0.00 & \cellcolor[rgb]{1.00, 1.00, 1.00} 0.00 & \cellcolor[rgb]{1.00, 1.00, 1.00} 0.00 & \cellcolor[rgb]{1.00, 1.00, 1.00} 0.00 &      &      & \cellcolor[rgb]{1.00, 1.00, 1.00} 0.00 & \cellcolor[rgb]{1.00, 1.00, 1.00} 0.00 & \cellcolor[rgb]{1.00, 1.00, 1.00} 0.00 &      & \cellcolor[rgb]{1.00, 1.00, 1.00} 0.00 & \cellcolor[rgb]{1.00, 1.00, 1.00} 0.00 & \cellcolor[rgb]{1.00, 1.00, 1.00} 0.00 & \cellcolor[rgb]{1.00, 1.00, 1.00} 0.00 & \cellcolor[rgb]{1.00, 1.00, 1.00} 0.00 & \cellcolor[rgb]{1.00, 1.00, 1.00} 0.00 & \cellcolor[rgb]{1.00, 1.00, 1.00} 0.00 & \cellcolor[rgb]{1.00, 1.00, 1.00} 0.00 & \cellcolor[rgb]{1.00, 1.00, 1.00} 0.00 & \cellcolor[rgb]{1.00, 1.00, 1.00} 0.00 & \cellcolor[rgb]{1.00, 1.00, 1.00} 0.00 & \cellcolor[rgb]{1.00, 1.00, 1.00} 0.00\\ 
v4 & \cellcolor[rgb]{0.98, 0.98, 0.98} 0.53 & \cellcolor[rgb]{0.98, 0.98, 0.98} 0.53 & \cellcolor[rgb]{0.98, 0.98, 0.98} 0.47 & \cellcolor[rgb]{1.00, 1.00, 1.00} 0.08 & \cellcolor[rgb]{1.00, 1.00, 1.00} 0.00 & \cellcolor[rgb]{0.98, 0.98, 0.98} 0.51 & \cellcolor[rgb]{0.99, 0.99, 0.99} 0.42 &      &      & \cellcolor[rgb]{0.98, 0.98, 0.98} 0.53 & \cellcolor[rgb]{0.98, 0.98, 0.98} 0.49 & \cellcolor[rgb]{1.00, 1.00, 1.00} 0.10 &      & \cellcolor[rgb]{0.98, 0.98, 0.98} 0.53 & \cellcolor[rgb]{0.98, 0.98, 0.98} 0.53 & \cellcolor[rgb]{1.00, 1.00, 1.00} 0.08 & \cellcolor[rgb]{1.00, 1.00, 1.00} 0.00 & \cellcolor[rgb]{0.98, 0.98, 0.98} 0.52 & \cellcolor[rgb]{1.00, 1.00, 1.00} 0.12 & \cellcolor[rgb]{1.00, 1.00, 1.00} 0.00 & \cellcolor[rgb]{1.00, 1.00, 1.00} 0.00 & \cellcolor[rgb]{0.98, 0.98, 0.98} 0.52 & \cellcolor[rgb]{1.00, 1.00, 1.00} 0.11 & \cellcolor[rgb]{1.00, 1.00, 1.00} 0.01 & \cellcolor[rgb]{1.00, 1.00, 1.00} 0.00\\ 
v5 & \cellcolor[rgb]{1.00, 1.00, 1.00} 0.04 & \cellcolor[rgb]{1.00, 1.00, 1.00} 0.04 & \cellcolor[rgb]{1.00, 1.00, 1.00} 0.03 & \cellcolor[rgb]{1.00, 1.00, 1.00} 0.01 & \cellcolor[rgb]{1.00, 1.00, 1.00} 0.00 & \cellcolor[rgb]{1.00, 1.00, 1.00} 0.04 & \cellcolor[rgb]{1.00, 1.00, 1.00} 0.03 &      &      & \cellcolor[rgb]{1.00, 1.00, 1.00} 0.04 & \cellcolor[rgb]{1.00, 1.00, 1.00} 0.03 & \cellcolor[rgb]{1.00, 1.00, 1.00} 0.01 &      & \cellcolor[rgb]{1.00, 1.00, 1.00} 0.04 & \cellcolor[rgb]{1.00, 1.00, 1.00} 0.03 & \cellcolor[rgb]{1.00, 1.00, 1.00} 0.01 & \cellcolor[rgb]{1.00, 1.00, 1.00} 0.00 & \cellcolor[rgb]{1.00, 1.00, 1.00} 0.03 & \cellcolor[rgb]{1.00, 1.00, 1.00} 0.03 & \cellcolor[rgb]{1.00, 1.00, 1.00} 0.01 & \cellcolor[rgb]{1.00, 1.00, 1.00} 0.00 & \cellcolor[rgb]{1.00, 1.00, 1.00} 0.03 & \cellcolor[rgb]{1.00, 1.00, 1.00} 0.03 & \cellcolor[rgb]{1.00, 1.00, 1.00} 0.01 & \cellcolor[rgb]{1.00, 1.00, 1.00} 0.00\\ 
v6 & \cellcolor[rgb]{0.70, 0.70, 0.70} 5.40 & \cellcolor[rgb]{0.71, 0.71, 0.71} 5.36 & \cellcolor[rgb]{0.72, 0.72, 0.72} 5.22 & \cellcolor[rgb]{0.81, 0.81, 0.81} 4.02 & \cellcolor[rgb]{0.89, 0.89, 0.89} 2.65 & \cellcolor[rgb]{0.72, 0.72, 0.72} 5.24 & \cellcolor[rgb]{0.73, 0.73, 0.73} 5.11 &      &      & \cellcolor[rgb]{0.70, 0.70, 0.70} 5.40 & \cellcolor[rgb]{0.72, 0.72, 0.72} 5.24 & \cellcolor[rgb]{0.77, 0.77, 0.77} 4.47 &      & \cellcolor[rgb]{0.69, 0.69, 0.69} 5.57 & \cellcolor[rgb]{0.71, 0.71, 0.71} 5.28 & \cellcolor[rgb]{0.79, 0.79, 0.79} 4.27 & \cellcolor[rgb]{0.86, 0.86, 0.86} 3.20 & \cellcolor[rgb]{0.72, 0.72, 0.72} 5.18 & \cellcolor[rgb]{0.76, 0.76, 0.76} 4.63 & \cellcolor[rgb]{0.87, 0.87, 0.87} 2.93 & \cellcolor[rgb]{0.96, 0.96, 0.96} 1.19 & \cellcolor[rgb]{0.73, 0.73, 0.73} 5.04 & \cellcolor[rgb]{0.78, 0.78, 0.78} 4.42 & \cellcolor[rgb]{0.90, 0.90, 0.90} 2.44 & \cellcolor[rgb]{0.97, 0.97, 0.97} 0.96\\ 
v7 & \cellcolor[rgb]{0.61, 0.61, 0.61} 6.42 & \cellcolor[rgb]{0.61, 0.61, 0.61} 6.38 & \cellcolor[rgb]{0.61, 0.61, 0.61} 6.39 & \cellcolor[rgb]{0.65, 0.65, 0.65} 5.92 & \cellcolor[rgb]{0.81, 0.81, 0.81} 3.99 & \cellcolor[rgb]{0.61, 0.61, 0.61} 6.38 & \cellcolor[rgb]{0.62, 0.62, 0.62} 6.22 &      &      & \cellcolor[rgb]{0.61, 0.61, 0.61} 6.40 & \cellcolor[rgb]{0.61, 0.61, 0.61} 6.44 & \cellcolor[rgb]{0.66, 0.66, 0.66} 5.88 &      & \cellcolor[rgb]{0.61, 0.61, 0.61} 6.38 & \cellcolor[rgb]{0.59, 0.59, 0.59} 6.62 & \cellcolor[rgb]{0.65, 0.65, 0.65} 5.92 & \cellcolor[rgb]{0.77, 0.77, 0.77} 4.56 & \cellcolor[rgb]{0.61, 0.61, 0.61} 6.37 & \cellcolor[rgb]{0.68, 0.68, 0.68} 5.64 & \cellcolor[rgb]{0.86, 0.86, 0.86} 3.22 & \cellcolor[rgb]{0.95, 0.95, 0.95} 1.52 & \cellcolor[rgb]{0.61, 0.61, 0.61} 6.35 & \cellcolor[rgb]{0.71, 0.71, 0.71} 5.36 & \cellcolor[rgb]{0.91, 0.91, 0.91} 2.23 & \cellcolor[rgb]{0.98, 0.98, 0.98} 0.57\\ 
\bottomrule
\end{tabular}
}
\vspace{-5mm}
\end{table*}

A complete version of Table~\ref{tab:eer} is presented in Table~\ref{tab:app:eer}. The results using fine-tuning sets from \symboltag{v3}, \symboltag{v4}, and \symboltag{v5}, and the results on the test sets of the three versions are added. The TTS attacks in \symboltag{v2-5} of MLAAD are likely similar to each other. Hence, the EER and drift values are more or less similar across these subsets.